\documentstyle[11pt]{article}
\begin{document}
\begin{center}
{\large\bf
Effective Potential of $\lambda\phi^{4}_{1+3}$ at Zero and Finite
Temperature}
\end{center}
\vspace{.7cm}
\centerline{Guang-jiong Ni$^*$\footnotetext{$^*$E-mail:
gjni@fudan.ihep.ac.cn} and Su-qing Chen}
\vspace{0.2cm}
\centerline{Department of Physics,
Fudan University, Shanghai 200433, China}
\vspace{0.7cm}
\centerline{abstract}
The effective potential of $\lambda\phi^4_{1+3}$ model with both sign of
parameter $m^2$ is evaluated at $T=0$ by means of a simple but effective
method for regularization and renormalization. Then at $T\ne 0$, the
effective potential is evaluated in imaginary time Green Function approach,
using the Plana formula. A critical temperature for restoration of
symmetry breaking in the standard model of particle physics is estimated
to be $T_c\simeq 510$ GeV.
\vspace*{5mm}
 
\newpage

\section{introduction}

The symmetry breaking (SB) in quantum field theory (QFT) is an important
problem both for particle physics and cosmology. Among various field
models, the $\lambda\phi^4_{1+3}$ model attracts much attention. Its
Lagrangian reads
\begin{equation}
{\cal L}\{\phi(x)\}=\frac{1}{2}\partial_{\mu}\phi\partial^{\mu}\phi
-\frac{1}{2}m^2\phi^2-\frac{\lambda}{4!}\phi^4
\end{equation}
The coefficient of $\phi^2$ term may be positive ($m^2>0$) or negative
($m^2=-\sigma<0$, $\sigma>0$). In the latter case, the model has
spontaneous symmetry breaking (SSB) at the tree level, i.e., the lowest
ground state (vacuum) will be shifted from $\phi=0$ to
$\phi=(6\sigma/\lambda)^{1/2}$. If the SB is caused by high loop
($L\ge 1$) calculation (so called as quantum radiative correction), then
the SB is refered to as dynamical symmetry breaking (DSB). For studying
the SB, a systematic method of loop  expansion was developed by Coleman
and Weinberg [1]. An effective potential (EP) $V(\hat{\phi})$ is derived
to show where is the stable vacuum with $\hat{\phi}$ being the constant
configuration of $<\phi>$, which is the quantum average (at zero
temperature, $T=0$) or thermodynamic average (at finite temperature,
$T\ne 0$) of field $\phi (x)$. The theory of EP method was further
elaborated by Jackiw [2]. After the pioneering work of Kirzhnits and
Linde [3] and the suggestion of S. Weinberg, the EP at finite
temperature was investigated by various authors [4-6]. It was found
that SSB may be restored at some critical temperature $T_c$, i.e.,
$V(\hat{\phi})|_{T=T_c}$ takes a minimum at $\hat{\phi}=0$ again.

The aim of this paper is to restudy SB by some new method in
calculation. In Section II, a simple regularization and renormalization
method will be introduced so that the ultraviolet divergence in Feynman
diagram integral (FDI), the counter term and the ambiguity between bare
and renormalized parameters will disappear. Then in Sections 3, 4 and 5,
the EP at $T=0$ and $T\ne0$ will be evaluated respectively. For $T\ne 0$
case, basing on imaginary time Green Function method, the Plana Formula is
used. For $m^2<0$ case, a critical temperature for restoration of SSB in the
standard model is estimated to be 510 GeV. In Sec. 6 and 7, we consider
the high loop correction and the fermion contribution to EP if a
coupling term between  a fermion field $\psi(x)$ and $\phi(x)$ is added
to Eq. (1.1). The final Section 8 contains a summary and discussion. An
Appendix is added to explain the Plana Formula.

\section{mass correction -- self energy}

As a warm up, let us consider the simplest nontrivial FDI in
$\lambda\phi^4$ theory. The 1 loop contribution to mass comes from the
self energy diagram $\Sigma$ or tadpole diagram :
\begin{equation}
\Sigma = \frac{1}{2}\lambda \int\frac{d^4
k}{(2\pi)^4}\frac{i}{k^2-m^2+i\epsilon}
\end{equation}
with 1/2 being the symmetry factor. Turning to Euclidean space with the
subscript $E$:
\begin{equation}
\Sigma=\frac{1}{2}\lambda \int\frac{d^4
k_E}{(2\pi)^4}\frac{1}{k^2_E+m^2}~.
\end{equation}
this FDI is quadratically divergent. Various kinds of regularization method
have been proposed for handling it. After learning all of them,
especially that in Refs. [7-11], begining from J-f Yang, we proposed a simple 
trick for calculating the chiral anomaly  [12], i.e., to
differentiate the FDI with respect to the external momentum enough times
so that it becomes convergent. Then after integrating back to original
one, we got some arbitrary constants as the substitution of the original
divergence. We will extend this method to study the problem here with
the external momentum replaced by some mass parameter.

Hence we differentiate Eq. (2.2) with respect to $m^2$ twice,
\begin{eqnarray}
\frac{d^2\Sigma}{d (m^2)^2}&=&\lambda\int \frac{d^4
k_E}{(2\pi)^4}\frac{1}{(k^2_E+m^2)^3}\nonumber\\
&=&\frac{\lambda}{32\pi^2m^2}~.
\end{eqnarray}
Then we obtain after integrating two times:
\begin{equation}
\Sigma=\frac{\lambda}{32\pi^2}(m^2 \ln m^2+C m^2+C')
\end{equation}
where $C$ and $C'$ are two arbitrary constants. As we will see
immediately below that endowing $C$ and $C'$ some values amounts to some
renormalization and corresponding explanation of the parameter $m$. By
using the well-known chain approximation, the propagator is modified
from the lowest order one, $G_2$:
\begin{equation}
G_2(p)=\frac{i}{p^2-m^2}
\end{equation}
to
\begin{eqnarray}
\tilde{G_2}(p)&=&G_2+G_2(-i\Sigma)G_2+...\nonumber\\
&=&\frac{G_2}{1+i\Sigma G_2}=\frac{i}{p^2-m^2_R}
\end{eqnarray}
with
\begin{equation}
m^2_R=m^2+\Sigma=m^2+\frac{\lambda}{32\pi^2}(m^2\ln
\frac{m^2}{\mu^2}+C')
\end{equation}
where we rewrite the constant $C$ as $(-\ln\mu^2)$ with $\mu$ being
an arbitrary mass scale.

Now our renormalization procedure amounts to setting $C'=0$ so that when
$m^2=0$, $m^2_R=0$ also.
Furthermore
\begin{equation}
m^2_R|_{\mu=m}=m^2
\end{equation}
defines the observed mass of particle in free motion. It also gives the
explanation of parameter $m$ in the model (1.1). Note that, however, we
need not to change the notation of $m$ whether it is used at tree level
calculation  or in QFT at higher loop order as in Eq. (2.7). In the
latter case, the quantum field effect has been absorbed into the
definition of $m^2$.

It is at the very most we can do in performing a calculation on
self-energy $\Sigma$. And it is also the best way to express the fact
that we cannot obtain full information about mass generation mechanism
via simple calculation on some FDI. Let us think the converse, suppose
one finds a definite and finite $\delta m$ via some calculation on
$\Sigma$. then it would be possible to generate mass $m$ from a ``bare''
mass $m_0$ and the latter could be approaching to zero. Eventually one
would claim that he can generate a mass from a massless model. It is
incredible. Because if one generates a mass, say 3 grams, he must
generates a standard weight 1 gram at the same time. Only the dimensionless
ratio 3 can be calculated from a massless model . In short, what can
emerge from a massless model is either no mass scale or two mass scales,
but never one mass scale. This scenerio was clearly shown in the
Gross-Neveu model [13] and a reinvestigation on NJL model [14] in Ref. [15]
(see also [16]). The emergence of a massive fermion is accompanied by
the change (phase transition) of vacuum which provides another mass
scale (a standard weight). To a large extent, the stability and mass of
a particle is ensured by its environment and not merely by itself. (The
stability is related to the ``imaginary part'' of mass, i.e., the decay
constant. For instance, a neutron has different half-life time in different
nuclear environment).

So the mass generation mechanism is a nonperturbative process
accompanying the change of environment (vacuum). On the other hand, the
perturbative FDI calculation can at most provide some information about
the change of mass with some parameter. In the example here, Eq. (2.3)
does tell us the knowledge about $\frac{d^2\Sigma}{d(m^2)^2}$, but the FDI
cannot tell us definitely about $\Sigma$ or $\frac{d\Sigma}{d (m^2)}$.
Thus we understand that the emergence of divergence in FDI is
essentially a warning. It warns us that we expected too much. We
understand now that it is the indefiniteness or arbitrariness rather than
the divergence which is implicitly implied by the appearance of cut off
$\Lambda$ or pole $1/\epsilon$ in previous regularization methods.

Basing on above consideration, we are pleased to get rid of the
divergence, the counter term and the bare parameter.

\section{effective potential at zero temperature}

For comparison, let us use the formulas and notation in Ref. [5] while
denote $\hat{\phi}=\phi$ for brevity. The EP at tree level ($L=0$) reads
\begin{equation}
V_0(\phi)=\frac{1}{2}m^2\phi^2+\frac{\lambda}{4!}\phi^4
\end{equation}
  
The one-loop contribution to EP is evaluated [1,2,4-6] as
\begin{equation}
V_1(\phi)=\frac{1}{2}\int\frac{d^4 k_E}{(2\pi)^4}\ln
(k^2_E+m^2+\frac{1}{2}\lambda\phi^2)
\end{equation}

Denoting
\begin{equation}
M^2=m^2+\frac{1}{2}\lambda\phi^2
\end{equation}
we differentiate $V_1$ with respect to $M^2$ three times until it is
convergent
\begin{eqnarray}
\frac{\partial^3 V_1}{\partial (M^2)^3}&=&\int\frac{d^4
k_E}{(2\pi^4)^4}\frac{1}{(k^2_E+M^2)^3}\nonumber\\
=\frac{1}{2(4\pi)^2 M^2}
\end{eqnarray}
Then integrate it with respect to $M^2$ three times with the result
\begin{equation}
V_1=\frac{1}{2(4\pi)^2}\{\frac{M^4}{2}(\ln
M^2-\frac{1}{2})-\frac{1}{2}M^4+\frac{1}{2}C_1M^4+C_2 M^2+C_3\}
\end{equation}
where three arbitrary constants $C_1$, $C_2$ and $C_3$ are introduced.
The renormalization simply amounts to fix these constants at our
disposal.

For example, if $m^2=0$, we take $C_1=-\ln \mu^2$ with $\mu$ an arbitrary
mass scale and $C_2=C_3=0$. Eq. (3.5) is simplified to
\begin{equation}
V_1=\frac{1}{256\pi^2}[\lambda^2\phi^4\ln\frac{\lambda\phi^2}{2\mu^2}-\frac{3
\lambda ^2}{2}\phi ^4]
\end{equation}
Combining $V$ with $V_0$ in Eq. (3.1) and imposing the renormalization
condition
\begin{equation}
[\frac{d^4}{d\phi^4}(V_0+V_1)]_{\phi=M'}=\lambda
\end{equation}
one recovers the famous Coleman-Weinberg EP [1]:
\begin{equation}
V(\phi)=\frac{\lambda}{4!}\phi^4+\frac{\lambda^2\phi^4}{256\pi^2}(\ln
\frac{\phi^2}{M'^2}-\frac{25}{6})
\end{equation}

Now consider $m^2>0$ case. Choosing $C_1=-\ln m^2$, one has
\begin{equation}
V_1=\frac{1}{2(4\pi)^2}\{(m^2+\frac{1}{2}\lambda\phi^2)^2[\frac{1}{2}\ln\frac{(m^2
+
\lambda\phi^2/2)}{m^2}-\frac{3}{4}]+C_2(m^2+\frac{1}{2}\lambda\phi^2)+C_3\}
\end{equation}
If we further choose $C_2=m^2$, $C_3=-\frac{1}{4}m^4$, then
\begin{equation}
V_1|_{\phi=0}=\frac{\partial V_1}{\partial \phi}|_{\phi=0}
=\frac{\partial^2 V_1}{\partial \phi^2}|_{\phi=0}
=\frac{\partial^3 V_1}{\partial \phi^3}|_{\phi=0}
=\frac{\partial^4 V_1}{\partial \phi^4}|_{\phi=0}=0
\end{equation}
The whole EP combining $V_0$ and $V_1$ reads
\begin{equation}
V(\phi)=\frac{1}{2}m^2\phi^2+\frac{1}{4!}\lambda\phi^4
+\frac{1}{2(4\pi)^2}\{(m^2+\frac{1}{2}\lambda\phi^2)^2[\frac{1}{2}\ln\frac{(m^2+\l
ambda\phi^2/2)}{m^2}-\frac{3}{4}]
+m^2(\frac{1}{2}\lambda\phi^2+\frac{3}{4}m^2)\}
\end{equation}
with
\begin{equation}
m^2=\frac{\partial^2 V}{\partial \phi^2}|_{\phi=0},~~
\lambda=\frac{\partial^4 V}{\partial \phi^4}|_{\phi=0},
\end{equation}

For $m^2=-\sigma<0$ (SSB case), the expression (3.5) remains valid with
$M^2=-\sigma+\frac{1}{2}\lambda\phi^2$. We take $C_1=-\ln\mu^2$, then
\begin{equation}
V=-\frac{1}{2}\sigma\phi^2+\frac{\lambda}{24}\phi^4+\frac{1}{2(4\pi)^2}\{
(\frac{1}{2}\lambda\phi^2-\sigma)^2[\frac{1}{2}\ln(\frac{\lambda\phi^2/2-\sigma}{\
mu^2})
-\frac{3}{4}]+C_2(\frac{1}{2}\lambda\phi^2-\sigma)+C_3\}
\end{equation}
and
\begin{equation}
\frac{dV}{d\phi}=\phi\{-\sigma+\frac{\lambda}{6}\phi^2+\frac{\lambda}{2(4\pi)^2}
[(\frac{1}{2}\lambda\phi^2-\sigma)(\ln\frac{\lambda\phi^2/2-\sigma}{\mu^2}-1)
+C_2]\}=0
\end{equation}
gives two solutions. One is
\begin{equation}
\phi_0=0~~~~~{(\rm symmetric~ phase)}
\end{equation}
To render the another one
\begin{equation}
\phi_1^2=6\sigma/\lambda~~~~~~~({\rm SSB~ phase})
\end{equation}
formally coinciding with that from tree level. The choice of
$\mu^2=C_2=2\sigma$ is necessary which also makes the mass excited at
the broken vacuum has the same expression as that at the tree level:
\begin{equation}
m^2_{\sigma}=\frac{d^2V}{d\phi^2}|_{\phi=\phi_1}=2\sigma
\end{equation}

The derivatives of different order at two phases are summarized in the
Table 1. The above assignment of $C_1$, $C_2$ and $C_3$ renders the
appearance of imaginary part in $V$ and its derivatives at symmetric
phase ($\phi_0=0$). It means the unstability of symmetric phase at the
presence of stable SSB phase.

It is interesting to see that another choice
\begin{equation}
C_1=-\ln(-\sigma),~C_2=-\sigma,~C_3=-\frac{1}{4}\sigma^2
\end{equation}
leaves only $\phi_0=0$ an extremum, i.e., semistable state with
\begin{equation}
V(0)=\frac{dV}{d\phi}|_{\phi=0}=\frac{d^3V}{d\phi^3}|_{\phi=0}=0
\end{equation}
and
\begin{equation}
\frac{d^2 V}{d\phi^2}|_{\phi=0}=-\sigma,~~\frac{d^4
V}{d\phi^4}|_{\phi=0}=\lambda
\end{equation}
No real SSB solution $\phi^2_1\ne 0$ exists. Hence we see that two
different choices of $C_i$ lead to two different sectors of EP. The
implication will be discussed at the final section.

Actually, we can
always perform the high loop renormalization such that Eqs. (3.16) and
(3.17) remain valid at any high order. This will be quite beneficial to
discuss the recovery of SSB at high temperature (see section 5).

\section{Effective potential at finite temperature}

As shown clearly in Refs. [4-6],
to study the temperature effects in QFT, one of most widely used methods
is the imaginary time Green function approach, which amounts to replace
the continuous fourth (Euclidean) momentum $k_4$ by discrete $\omega_n$
and integration by a summation ($\beta=1/k_B T$):
\begin{eqnarray}
k_4&\to& \omega_n=\left\{
\begin{array}{ll}
\frac{2\pi n}{\beta}~, & n=0,\pm 1,.. ({\rm boson})\\
\frac{\pi(2n+1)}{\beta}~, & n=0,\pm 1, ... ({\rm fermion})
\end{array}
\right.\nonumber\\
\int\frac{d^4 k}{(2\pi)^4} &\to&
\sum_n\frac{1}{\beta}\int\frac{d^3k}{(2\pi)^3}
\end{eqnarray}
Thus instead of Eq. (3.2), the one-loop contribution to EP at $T\ne 0$
reads
\begin{equation}
V_1^{\beta}(\phi)=\frac{1}{2\beta}\sum_n
\int\frac{d^3k}{(2\pi)^3}\ln[{\rm\bf k}^2+(\frac{2\pi
n}{\beta})^2+m^2+\frac{1}{2}\lambda\phi^2]
\end{equation}
Let us evaluate $V^{\beta}_1$ by a rigorous trick. Because
$V^{\infty}_1(\phi)\equiv V_1(\phi)$ is already known, what we need is
the difference, $V^{\beta}_1-V_1$, which turns out to be finite:
\begin{eqnarray}
V^{\beta}_1(\phi)-V_1(\phi)&=&\frac{1}{2\beta}
\int\frac{d^3k}{(2\pi)^3}(\sum_n-\int dn)f(n)\nonumber\\
f(n)&=&\ln(n^2+b^2)\nonumber\\
b^2&=&\frac{\beta^2}{4\pi^2}({\rm\bf k}^2+m^2+\frac{1}{2}\lambda\phi^2)
\end{eqnarray}
Since $(\sum_{n=-\infty}^{\infty}-\int_{n=-\infty}^{\infty} dn)f(n)=
2({\sum_{n=0}^{\infty}}'-\int_{0}^{\infty}dn)f(n)$, where
${\sum_{n=0}^{\infty}}'f(n)=\frac{1}{2}f(0)+\sum_{n=1}^{\infty}f(n)$, we manage
to calculate it by Plana formula [16,17] (see Appendix). Note that
$f(z)$ has two branch points at $z=-bi$ and $bi$. Introducing a cut from
$-bi$ to $bi$, denoting $arg(z+bi)=\theta_1>0$ (anticlock wise),
$arg(z-bi)=\theta_2<0$ (clockwise) and taking $\theta_1=\theta_2=0$ at
the left side of cut, one has ($z=it$):
\begin{equation}
f(z)=\left\{
\begin{array}{cc}
\ln|b-t|+\ln|-b-t|-i\pi~, & (-\infty<t<-b)\\
\ln|b+t|+\ln|b-t|~, & (-b<t<b)\\
\ln|b+t|+\ln|t-b|+i\pi~, & (b<t<\infty)
\end{array}
\right.
\end{equation}
Hence the fomula (A.1) can be used to yield
\begin{equation}
({\sum_{n=0}^{\infty}}'-\int_{0}^{\infty}dn)f(n)=-2\pi\int_b^{\infty}\frac{dt}{e^{
2\pi
t}-1}
\end{equation}
Substituting (4.5) into (4.3) and changing the order of integration with
respect to {\bf k} and $t$, one has
\begin{eqnarray}
V^{\beta}_1(\phi)-V_1(\phi)&=&-\frac{1}{\pi\beta}\int_{\frac{\beta}{2\pi}\sqrt{m^2
+\lambda\phi^2/2}}^{\infty}
\frac{dt}{e^{2\pi t}-1}\int_0^{[(\frac{2\pi
t}{\beta})^2-m^2-\frac{1}{2}\lambda\phi^2]^{1/2}} k^2 dk\nonumber\\
&=&-\frac{8\pi^2}{3\beta^4}\int_{\frac{\beta}{2\pi}\sqrt{m^2+\lambda\phi^2/2}}^{\i
nfty}
\frac{t^3 dt}{e^{2\pi t}-1}(1-\frac{m^2+\lambda\phi^2/2}{(2\pi
t/\beta)^2})^{3/2}
\end{eqnarray}

Eq. (4.6) is rigorous but difficult to evaluate.  As a high temperature
approximation of (4.6), $T\to\infty$, $\beta\to 0$, we expand the
parentheses to second term and set the low limit of integral as zero to
yield:
\begin{eqnarray}
V^{\beta}_1(\phi)-V_1(\phi)&\simeq&
-\frac{8\pi^2}{3\beta^4}\int_0^{\infty}\frac{t^3dt}{e^{2\pi t}-1}+
\frac{m^2+\lambda\phi^2/2}{\beta^2}\int_0^{\infty}\frac{t dt}{e^{2\pi
t}-1}\nonumber\\
&=&-\frac{\pi^2}{90\beta^4}+\frac{m^2+\lambda\phi^2/2}{24\beta^2},~~
(T\to\infty, \beta\to 0)
\end{eqnarray}
which coincides with Eq. (3.16) in Ref. [5] to the second term.

\section{Symmetry restoration at finite temperature}

In this case the EP at $T=0$ is given by (3.13). For evaluting the EP at
$T>0$, we cannot use the results in Sec. 4 directly for small $\phi$.
Because now in Eq. (4.3) $m^2=-\sigma<0$.
\begin{equation}
b^2=-\frac{\beta^2}{4\pi^2}[(\sigma-\frac{1}{2}\lambda\phi^2)-{\bf\rm
k}^2]\equiv -a^2<0~, ~~~({\rm if } \lambda\phi^2/2<\sigma, {\bf\rm
k}^2<\sigma-\lambda\phi^2/2)
\end{equation}
So we have to divide the integration with respect to $k$ into two
regions
\begin{eqnarray}
V^{\beta}_1(\phi)-V_1(\phi)&=&\frac{1}{4\pi^2\beta}(I_1+I_2)\nonumber\\
I_1&=&\int_0^{(\sigma-\lambda\phi^2/2)^{1/2}} k^2 dk
2({\sum_{n=0}^{\infty}}'-\int_{0}^{\infty}dn)
\ln(n^2-a^2)\nonumber\\
I_2&=&\int_{(\sigma-\lambda\phi^2/2)^{1/2}}^{\infty} k^2 dk
2({\sum_{n=0}^{\infty}}'-\int_{0}^{\infty}dn)
\ln(n^2+b^2)
\end{eqnarray}
Evidently, $I_2$ can be calculated as that in Sec. 4. Let us concentrate
on $I_1$.

Fortunately, there is a formula derived by Barton [18]
\begin{equation}
({\sum_{n=0}^{\infty}}'-\int_{0}^{\infty}dn)\ln|\eta+n|=-2\int_0^{\infty}
\frac{dt}{e^{2\pi
t}-1}\tan^{-1}\frac{t}{\eta}+\theta(-\eta)\ln|2\sin\pi\eta|
\end{equation}
The first term changes sign with that of $\eta$, so
\begin{equation}
({\sum_{n=0}^{\infty}}'-\int_{0}^{\infty}dn)\ln(n^2-a^2)
=\ln|2\sin\pi a|
\end{equation}
\begin{equation}
I_1=2\int_0^{(\sigma-\lambda\phi^2/2)^{1/2}}
k^2dk\ln[2\sin\frac{\beta}{2}\sqrt{\sigma-\lambda\phi^2/2-k^2}]
\end{equation}
\begin{equation}
I_2=-4\pi\int_0^{\infty}\frac{dt}{e^{2\pi t}-1}
\int_{(\sigma-\lambda\phi^2/2)^{1/2}}^{[(2\pi
t/\beta)^2+\sigma-\lambda\phi^2/2]^{1/2}} k^2 dk
\end{equation}
In (5.3), after integration with respect to $k$, we can expand it with
respect to small $t$ and keep one term only because the integration is
dominant there
\begin{eqnarray}
I_2&\simeq&-4\pi\frac{1}{3}\int_0^{\infty}\frac{dt}{e^{2\pi t}-1}
(\sigma-\frac{1}{2}\lambda\phi^2)^{3/2}[\frac{3}{2}\frac{4\pi^2
t^2/\beta^2}{\sigma-\lambda\phi^2/2}]\nonumber\\
&=&-\frac{8\pi^3}{\beta^2}(\sigma-\frac{1}{2}\lambda\phi^2)^{1/2}
\int_0^{\infty}\frac{t^2 dt}{e^{2\pi t}-1}
=-\frac{2\zeta(3)}{\beta^2}(\sigma-\frac{1}{2}\lambda\phi^2)^{1/2}
\end{eqnarray}
where $\zeta(3)=1.202$.

We are interested in finding the critical temperature $\beta_c$ for
restoring the SSB. There are two possible points of view:

(a) Following Ref. [5] and noting Eq. (3.20) or Table 1, we may
determine $\beta_c$ from the following condition
\begin{equation}
[\frac{\partial^2}{\partial\phi^2}(V^{\beta}_1-V_1)]_{\phi=0}=\sigma
\end{equation}
which means that the curvature of EP at $\phi=0$,
$[\frac{\partial^2}{\partial \phi^2} V^{\beta}_1]_{\phi=0}$, turn from
negative to positive at $T_c$.

Combining Eqs. (5.2)-(5.7), we obtain after an integration by part:
\begin{equation}
[\frac{\partial^2}{\partial \phi^2} (V^{\beta}_1-V_1)]_{\phi=0}
=\frac{1}{4\pi^2\beta}\{-\lambda\int_0^{\sqrt{\sigma}} dk
\ln[2\sin(\frac{\beta}{2}\sqrt{\sigma-k^2})]+\frac{\zeta(3)}{\beta^2}\frac{\lambda
}{\sigma^{1/2}}\}
\end{equation}
In the high temperature approximation, the first term can be neglected:
\begin{equation}
[\frac{\partial^2}{\partial \phi^2} (V^{\beta}_1-V_1)]_{\phi=0}
=\frac{\zeta(3)}{4\pi^2\beta^3}\frac{\lambda}{\sigma^{1/2}},
~~(T\to\infty,\beta\to 0)
\end{equation}
Combining (5.8) and (5.10), we find the critical temperature
\begin{equation}
\frac{1}{\beta_c}=[\frac{4\pi^2}{\lambda\zeta(3)}]^{1/3}\sigma^{1/2}\simeq
3.202\frac{\sigma^{1/2}}{\lambda^{1/3}}~,
(\beta\to 0)
\end{equation}

The result of Eq. (3.17) in Ref. [5] reads
\begin{equation}
\frac{1}{\beta_c}=\sqrt{\frac{24}{\lambda}\sigma}
=4.899\frac{\sigma^{1/2}}{\lambda^{1/2}}~,
(\beta\to 0)
\end{equation}
which was derived formally from the expression valid for $m^2>0$ as can
also be seen from Eq. (4.7) in this paper.

(b)But for the restoration of SSB, a better criterion should be found
directly from the vanishing of curvature at $\phi=\phi_1$. Basing on
(4.7) and (3.13), we find the broken vacuum is shifted to
$\phi^{\beta}_1=\sqrt{\frac{6\sigma^{\beta}}{\lambda}}$ with approximately
\begin{equation}
\sigma^{\beta}=\sigma-\frac{\lambda}{24\beta^2}
\end{equation}
$\mu^2=C_2=2\sigma$ as before. Then to the order $O(\lambda)$,
the vanishing condition of
\begin{equation}
\frac{d^2V^{\beta}_1}{d\phi^2}|_{\phi=\phi^{\beta}_1}=2\sigma-\frac{\lambda}{12
\beta ^2}
\end{equation}
yields the critical temperature
\begin{equation}
T_c=\frac{1}{\beta_c}=\sqrt{\frac{12}{\lambda}}m_{\sigma}
\end{equation}
$m_{\sigma}$ being the excitation mass at $T=0$. It is pleased to see
that Eq. (5.15) is conciding with (5.12) (i.e., (3.20) in Ref. [5]).

As an interesting application of (5.15), in the standard model
(electro-weak unified theory) of particle physics, the mass square of
Higgs particle reads at tree level as
\begin{equation}
m^2_H=\frac{2}{3}\lambda|<\phi>|^2
\end{equation}
while $|<\phi>|$ can be evaluated from the mass of $W$ boson
\begin{equation}
m_W=\frac{1}{\sqrt{2}}g|<\phi>|
\end{equation}
to be of the order
\begin{equation}
|<\phi>|\simeq 180 GeV
\end{equation}
combination of (5.15) ($m_{\sigma}\to m_H$) with (5.16) and (5.18) leads
to an estimation that the SSB of standard model would be restored at a
critical temperature
\begin{equation}
T_c=\sqrt{8}|<\phi>|\simeq 510 GeV
\end{equation}

\section{high loop correction}

Let us consider the higher loop correction, for example, the dominant
two-loop contribution at $T=0$ in a $O(N)$ $\lambda\phi^4$ model for
large $N$ reads (see Eq. (3.28a) in Ref. [5]):
\begin{eqnarray}
V_2(\phi)&=&\frac{1}{6}\lambda(\frac{1}{2}NI)^2\nonumber\\
I&=&\int \frac{d^4k}{(2\pi)^4}\frac{1}{k^2+m^2+\lambda\phi^2/2}
\end{eqnarray}
Using the same regularization-renormalization trick as that in Sec. 3,
we arrive at
\begin{equation}
V_2(\phi)=\frac{\lambda
N^2}{24}\frac{1}{(4\pi)^4}\{(m^2+\frac{1}{2}\lambda\phi^2)[\ln
\frac{(m^2+\lambda\phi^2/2)}{\mu^2}-1]+C_2\}^2
\end{equation}
One can set $\mu^2=m^2$ and $C_2=m^2$ such that
\begin{equation}
V_2(\phi)|_{\phi^2=0}=\frac{d V_2}{d\phi^2}|_{\phi^2=0}=0
\end{equation}
as mentioned at the end of Sec. 3. For $m^2=-\sigma<0$ case, similar
formula can be found for SSB occurs at $\phi^2_1=2\sigma$
($\mu^2=C_2=2\sigma$). Furthermore, we consider the
correction at $T\ne 0$
\begin{eqnarray}
V_2^{\beta}-V_2&=&\frac{\lambda
N^2}{24}\{[\int\frac{d^3k}{(2\pi)^3}\frac{1}{\beta}\sum_{-\infty}^{\infty}
f(n)]^2-[\int
\frac{d^3k}{(2\pi)^3}\frac{1}{\beta}\int_{-\infty}^{\infty}dn
f(n)]^2\}\nonumber\\
f(n)&=&\frac{\beta^2}{4\pi^2 [n^2+\frac{\beta^2}{4\pi^2}({\bf\rm
k}^2+m^2+\lambda\phi^2/2)]}
\end{eqnarray}
Note that $f(n)$ is now a single-valued even function of $n$, so
\begin{equation}
V_2^{\beta}-V_2=\frac{\lambda
N^2}{24}\{\int\frac{d^3k}{(2\pi)^3}\frac{1}{\beta}(\sum_{n=-\infty}^{\infty}
+\int_{-\infty}^{\infty}dn)f(n)\}\{\int
\frac{d^3k}{(2\pi)^3}\frac{2}{\beta}({\sum_{n=0}^{\infty}}'
-\int_{0}^{\infty}dn)f(n)\}=0
\end{equation}
due to Eq. (A.2).

We see that in this example there is no temperature dependent correction at 
two-loop
order. It seems to us that this property persists to any higher order.
Therefore, all temperature dependent correction to EP comes from
one-loop graph as shown in Eq. (4.6)-(4.7) for $m^2>0$ or Eq.
(5.2)-(5.7) for $m^2<0$, at least for $O(N) \lambda \phi^4$ model at large 
$N$ limit.

\section{contribution of fermion field to eeffective potential}
Assuming that there is a fermion field $\psi(x)$ coupling to the
multiplet of Bose fields $\phi_a(x)$ (see Eq. (4.6) in Ref. [5]):
\begin{equation}
{\cal L}\{\phi_a(x), \psi(x)\}=i\bar{\psi}\gamma^{\mu}\partial_{\mu}\psi
-m_f\bar{\psi}\psi-\bar{\psi}G^a\psi\phi_a+{\rm boson~ terms}
\end{equation}
then the shifted ``free'' Lagrangan is
\begin{equation}
\hat{{\cal L}_0}\{\hat{\phi},\phi(x), 
\psi(x)\}=i\bar{\psi}\gamma^{\mu}\partial_{\mu}\psi
-\bar{\psi}M\psi+{\rm boson~ terms}
\end{equation}

If the mass matrix after SB at $<\phi_a>=\hat{\phi}_a$
\begin{equation}
M=m_f+G^a\hat{\phi}_a
\end{equation}
has the $i$th eigenvalue $M_i$, then one-loop EP at $T=0$, apart from
the boson contribution reads
\begin{equation}
V_{f1}(\hat{\phi})=-2\sum_{i}\int\frac{d^4k_E}{(2\pi)^4}\ln(k^2_E+M^2_i)
\end{equation}
As in Sec. 3, we obtain
\begin{equation}
V_{f1}(\hat{\phi})=-2\frac{1}{(4\pi)^2}\sum_i\{\frac{1}{2}M^4_i\ln\frac{M^2_i}{\mu
^2}
-\frac{3}{4}M^4_i+C_2M^2_i+C_3\}
\end{equation}
For simplicity, we consider the case of one species: $M=m_f+G\phi_1$,
($\hat{\phi}=\phi$), and take $\mu^2=2\sigma=m^2_{\sigma}$, the mass
square of excitation at broken phase $\phi=\phi_1$
($\phi^2_1=6\sigma/\lambda$ in the absence of fermion). Then the
constant $C_2$ can be chosen so that
\begin{equation}
\frac{\partial V_{f1}}{\partial\phi}|_{\phi=\phi_1}=0
\end{equation}
which leads to
\begin{equation}
\frac{\partial^2 V_{f1}}{\partial
\phi^2}|_{\phi=\phi_1}=-\frac{8G^2}{(4\pi)^2}(m_f+G\phi_1)^2\ln\frac{(m_f+G\phi_1)
^2}{m^2_{\sigma}}
\end{equation}
Hence we see that the coupling between $\phi$ field and a fermion with
mass $M=m_f+G\phi_1$ may increase or decrease the mass of $\phi$ field
depending on the condition $M<m_{\sigma}$ or $M>m_{\sigma}$.

An interesting question is the following. If in (7.2) one has only the
free boson term
\begin{equation}
V_0=\frac{1}{2}m_0^2\phi^2
\end{equation}
can the DSB occur via the coupling with a fermion?

Suppose the DSB occurs at $\phi=\phi_1\ne 0$. For simplicity, we
consider the case of one species of fermion, so $M=m_f+G\phi_1$ in (7.5).
The criterion
\begin{equation}
[\frac{d}{d\phi}(V_0+V_{f1})]_{\phi=\phi_1}=0
\end{equation}
determines the constants
\begin{equation}
C_2=\frac{(4\pi)^2}{4G}\frac{m^2_0\phi_1}{(G\phi_1+m_f)}+(G\phi_1+m_f)^2
\end{equation}
with
\begin{equation}
\mu=G\phi_1+m_f
\end{equation}
chosen for convenience. Then the mass square of excitation at broken
vacuum reads
\begin{equation}
m^2_{\phi}=\frac{d^2V}{d\phi^2}|_{\phi_1}=m_0^2(1-\frac{G\phi_1}{G\phi_1+m_f})
\end{equation}
It looks meaningful. However, the value of $\phi_1$ in (7.10) and (7.11)
is autually arbitrary. This ambiguity reflects the fact that the DSB is
accompanying by the change of vacuum (environment), $\phi_1 \ne 0$,
which is far beyond the reach of perturbative QFT. Especially, if
$m_f=0$, $m^2_{\phi}=0$, irrespective of the value of $\phi_1$.

This situation does happen in NJL model [14], in the simpler version of
discrete symmetry
\begin{equation}
{\cal 
L}=i\bar{\psi}\gamma^{\mu}\partial_{\mu}\psi+\frac{1}{2}g^2(\bar{\psi}\psi)^2
\end{equation}
By adding a term $-\frac{1}{2}[m_0\phi+g(\bar{\psi}\psi)]^2$ with $\phi$
an auxiliary scalar field (with dimension $[M]$) and an arbitrary mass
scale $m_0$, we get
\begin{equation}
{\cal 
L}=i\bar{\psi}\gamma^{\mu}\partial_{\mu}\psi-\frac{1}{2}m^2_0\phi^2-G\phi\bar{\psi
}\psi
\end{equation}
where $G=gm_0$ is a dimensionless parameter. Then Eq. (7.12)tells us
that we cannot discuss the DSB of NJL model in the present scheme. An
improved nonperturbative treatment like Ref. [15] is needed.

It will be quite different to put the Lagrangian (7.13) in $1+1$
dimensional space, i.e., to discuss Gross-Neveu model [13]. By adding a
term $-\frac{1}{2}[\sigma+g(\bar{\psi}\psi)]^2$ with auxiliary field
$\sigma (\sim [M])$ and dimensionless $g$, we find the fermion
contribution at 1 loop level
\begin{eqnarray}
V_1(\sigma)&=&-\int\frac{d^2k_E}{(2\pi)^2}\ln(k^2_E+g^2\sigma^2)\nonumber\\
&=&\frac{1}{4\pi}[g^2\sigma^2(\ln\frac{g^2\sigma^2}{\mu^2}-1)+C_2]
\end{eqnarray}
with only two arbitrary constants $\mu^2$ and $C_2$. The DSB criterion
($V=V_0+V_1$):
\begin{equation}
[\frac{d}{d\sigma}V]|_{\sigma=\sigma_M}=0
\end{equation}
yields a relation between $\sigma_M$ and $\mu$
\begin{equation}
\sigma_M=\frac{\mu}{g}e^{-\pi/g^2}
\end{equation}
The mass square excited at $\sigma_M$ reads
\begin{equation}
m^2_{\sigma}=\frac{d^2
V}{d(\sigma/\sigma_M)^2}|_{\sigma_M}=\frac{g^2}{\pi}\sigma^2_M
\end{equation}

This is indeed meaningful as $m_{\sigma}$ and $\sigma_M$ are the only
two massive observables, now relating each other. Note also that in
(7.17), if $g^2\to 0^+$, $\sigma_M\to 0$, $g^2=0$ is an essential
singular point, so the mass generation mechanism is a nonperturbative
process.

After the excursion on DSB, let us return to the finite temperature
case. Using formula (4.1) and Eq. (7.4), we can easily write down the
temperature correction on 1 loop fermion contribution as
($E^2_{M_i}={\rm\bf k}^2+M^2_i$)
\begin{eqnarray}
V^{\beta}_{f1}-V_{f1}&=&-2\sum_i\int\frac{d^3 k}{(2\pi)^3}\frac{1}{\beta}
(\sum_{n=-\infty}^{\infty}-\int_{-\infty}^{\infty}dn)\ln[\frac{(2n+1)^2\pi^2}
{\beta ^2}
+E^2_{M_i}]\nonumber\\
&=&-\frac{2}{\beta}\sum_i\int\frac{d^3 k}{(2\pi)^3}
\{({\sum_{n=0}^{\infty}}'-\int_{0}^{\infty}dn)f(n)
+({\sum_{n=0}^{\infty}}'-\int_{0}^{\infty}dn)g(n)\}
\end{eqnarray}
where
\begin{eqnarray}
f(n)&=&\ln\frac{1}{4}[(2n+1)^2+4b^2]\nonumber\\
g(n)&=&\ln\frac{1}{4}[(2n-1)^2+4b^2]\nonumber\\
b^2=\frac{\beta^2E^2_{M_i}}{4\pi^2}
\end{eqnarray}
Similar to the trick used in Sec. 4, we first consider the multi-valued
function with complex variable $z$:
\begin{equation}
f(z)=\ln[(z+\frac{1}{2})^2+b^2]
\end{equation}
having two branch points
\begin{equation}
z_{2,1}=-\frac{1}{2}\pm ib
\end{equation}
Rewrite
\begin{equation}
f(z)=\ln(z-z_1)(z-z_2)\equiv \ln\rho_1\rho_2+i(\theta_1+\theta_2)
\end{equation}

Let $z=it$ running along the imaginary axis, ($-\infty<t<\infty$). Then
\begin{eqnarray}
\rho_1&=&[\frac{1}{4}+(t+b)^2]^{1/2},
~~\rho_2=[\frac{1}{4}+(t-b)^2]^{1/2}\nonumber\\
\theta_1&=&2\pi-\alpha_1,~~~~\alpha_1=\tan^{-1}\frac{1/2}{b+t}\nonumber\\
\theta_2&=&-(2\pi-\alpha_2),~~~~\alpha_2=\tan^{-1}\frac{1/2}{b-t}
\end{eqnarray}
Using Eq. (A.1), we obtain
\begin{equation}
({\sum}'-\int_0^{\infty}dn)f(n)=2\int_0^{\infty}\frac{dt}{e^{2\pi t}-1}
[\tan^{-1}\frac{1/2}{b+t}-\tan^{-1}\frac{1/2}{b-t}]
\end{equation}
 But it is just cancelled by the contribution of
$({\sum}'-\int_0^{\infty}dn)g(n)$, yielding
\begin{equation}
V^{\beta}_{f1}-V_{f1}=0
\end{equation}

So besides the dependence of $M_{i}$ via $\phi_{a}$ on the temperature,  
there is no extra temperature dependence in the fermion contribution to EP
of $\phi$ field, Eq. (7.5), at least at the 1 loop level.

\section{summary and discussion}

(a) Basing on previous experiences in literature, we propose to use a
simple but effective method for regularization and renormalization in
QFT. Once encountering a superficially divergent FDI, we first take its
derivative with respect to a mass parameter (or external momentum)
enough  times until it becomes convergent. After performing integration,
we reintegrate it with respect to the parameter the same times for
returning back to the original FDI. Now instead of divergence, several
arbitrary constants $C_i$ ($i=1,2,..$) appear in FDI. Further
renormalization amounts to reasonably fix $C_i$ at one's disposal .

(b) The physical essence of the appearance of these arbitrary constants
$C_i$, as discussed in Sec. 2, is ascribed to the lack of knowledge
about the model at the level of QFT under consideration.

To be precise, the $\lambda\phi^4$ model at classical field theory (CFT)
level is described by Lagrangian density ${\cal L}$, Eq. (1.1). While at QFT
level, it is described by effective Hamiltonian density:
\begin{equation}
{\cal H}_{\rm
eff}=\frac{1}{2}\dot{\phi}^2+\frac{1}{2}(\nabla\phi)^2+V_{\rm eff}(\phi)
\end{equation}
with EP $V_{\rm eff}$ given by $V_0+V_1+V_2+...$ in loop expansion as in
this paper.

The important thing lies in the fact that ${\cal H}_{\rm eff}$ is not
only different from ${\cal H}$ (derived from ${\cal L}$ at the classical
level) in form, but containing these arbitrary constants $C_1$, $C_2$
and $C_3$. While $C_3$ is trivial, the precise information, i.e., the
exact value of $C_i$ ($i=1,2$) is not contained in ${\cal L}$ and cannot
be predicted by the perturbative QFT. It is beyond our theoretical ability
and so can only be fixed by experiment via suitable renormalization
procedure.

(c) Similar situation occurs in the relation between quantum mechanics
(QM) and classical mechanics (CM). In CM, the motion of a particle is
determined by the classical Hamiltonian $H=T+V$. One can quantize it in
QM to solve the Schr\"{o}dinger Equation (SE)
\begin{equation}
\hat{H}\psi=i\hbar\frac{\partial\psi}{\partial t}
\end{equation}
with $\hat{H}=\hat{T}+V$ in configuration space. Usually $V$ does not
change. But in the case of singular potential, e.g.,
$V(r)=-Z\frac{e^2}{4\pi r}$, $Z>\frac{4\pi}{e^2}=137$ in Dirac Equation
or $V(r)\sim -\frac{1}{r^2}$ in SE, one soon runs into difficulty. After
many years research, (see Chapter 2 in Ref. [16]), people realize that
in this case the problem in QM is not well defined by Eq. (8.2), i.e.,
by $\hat{H}$ with suitable boundary conditions. We cannot find a
complete orthonormal set $\{\psi_i(\vec{x})\}$ unambiguously. What we can do
is to find a complete orthonormal set $\{\psi_i({\rm\bf x},\theta)\}$
which contains a (common) phase angle $\theta$. In other words, the
Hilbert space is divided into infinite sectors characterizing by a
continuous parameter $\theta$, which should be looked as a complement to
the usual problem in QM from outside. To fix $\theta$ is beyond the
ability of QM. An interesting example of a fermion moving in the field
of a dyon was discussed in Ref. [19] (see also [16]) with the value of
$\theta$ fixed eventually via the knowledge of QFT.

(d) Now for the QFT of $\lambda\phi^4$ model with $m^2>0$, since we are
staying at symmetric phase $\phi=0$, the renormalization (3.12)
determine $C_i$. As discussed in Sec. 2, this is the only reasonable
choice, because $m$ is the only observable mass parameter.

For the case of SSB, $m^2=-\sigma<0$, there are two observable mass
scales, the shifted vacuum, $\phi_1=\sqrt{\frac{6\sigma}{\lambda}}$, and
the excitation mass on it, $m_{\sigma}=\sqrt{2\sigma}$. They are related
(similar to that in Gross-Neveu model [13]) with $C_i$ shown
in Table 1. Note that, however, different choice of $C_i$ leads to
different sector of EP. Maybe it is just the counterpart of $\theta$ in
$\{\psi_i({\rm\bf x},\theta)\}$ of QM. Perturbative QFT is not well defined  by 
${\cal
L}$ given in CFT, it should be complemented by some $C_i$. In this sense,
$\lambda\phi^4$ model is nontrivial and renormalizable at QFT level.

(e) Due to the above advantage and the use of Plana formula, we evaluate
the EP of $\lambda\phi^4$ model at both zero and finite temperature with
the following results:

(i) At $T=0$, we manage to keep the location of stable vacuum and mass excitation
on it unaltered in form. Especially, in SSB case,
$\phi^2_1=6\sigma/\lambda$ and $m^2_{\sigma}=2\sigma$ remain as two
observed quantities up to any loop orders.

(ii) When $T>0$, the restoration of SSB ocuurs at some critical
temperature $T_c$ given by Eq. (5.15). A value of $T_c$ in the standard
model of particle physics is estimated to be $T_c\simeq 510$ GeV.

(iii) For pure $\lambda\phi^4$ model, the one-loop ($L=1$) contribution
to EP is temperature dependent. Whether higher ($L\ge 2$) contribution to EP
is temperature independent needs further study.

(iv) Further coupling with fermions may modify the EP of $\phi$ field,
but at least at $L=1$ order,their contributions seem also temperature independent.

\centerline{ACKNOWLEDGEMENTS}

We thank Dr. Ji-feng Yang for helpful discussions. This work was
supported in part by the National Science Foundation in China.

\appendix
\section{Plana Formula [17,16]}

{\bf Theorem}. Assuming a function $f(z)$ is analytic on the right half
complex plane, $Rez>0$, then
\begin{eqnarray}
\sum_{n=1}^{\infty}f(n)&+&\frac{1}{2}f(0)-\int_0^{\infty}f(x)dx\nonumber\\
&=&\int_{0 (C_2)}^{i\infty}\frac{f(z)dz}{e^{-2\pi iz}-1}
-\int_{-i\infty (C_1)}^{0}\frac{f(z)dz}{e^{2\pi iz}-1}
\end{eqnarray}
Here $C_1$ and $C_2$ are two integration contours along the right side
of imaginary axis on one sheet of Riemann surface of complex plane from
($-i\infty$) to 0 and from 0 to ($i\infty$) respectively.

If the integration along $C_1$ and $C_2$ does not encounter a cut, i.e.,
if $f(z)$ is a single-valued function, one may set $z=it$ in the
integrand directly and obtain:
\begin{eqnarray}
({\sum_n}'-\int_0^{\infty}dn)f(n)&\equiv&\sum_{n=1}^{\infty}f(n)+\frac{1}{2}f(0)
-\int_0^{\infty}f(n)dn\nonumber\\
&=&i\int_0^{\infty}\frac{f(it)-f(-it)}{e^{2\pi t}-1} dt
\end{eqnarray}
To our knowledge, some applications of Eq. (A.2) can be found in Refs.
[20,21] whereas Eq. (A.1) was used to derive the Casimir effect [17].
This Plana formula had also been used extensively by Barton [18].

\newpage
\begin{table}
  \caption{ $\mu^2=C_2=2\sigma$, 
    $C_3=-\sigma^2+(4\pi)^2\frac{3 \sigma^2}{\lambda}$}
\begin{center}
\begin{tabular}{ccc}
\hline
\hline
 & SSB phase & Symmetric phase\\
\hline
$\phi$ & $\phi_1=\sqrt{\frac{6\sigma}{\lambda}}$ & $\phi_0=0$ \\
\hline
$V$ & 0 & $-\frac{\sigma^2}{2(4\pi)^2}[\frac{15}{4}+\frac{1}{2}\ln2
-i\frac{\pi}{2}]+\frac{3}{2} \frac{\sigma^2}{\lambda}$\\
\hline
$\frac{dV}{d\phi}$ & 0 & 0\\
\hline
$\frac{d^2 V}{d\phi^2}$ & $2\sigma$ &
$-\sigma[1-\frac{\lambda}{2(4\pi)^2}(3+\ln 2-i \pi)]$\\
\hline
$\frac{d^3 V}{d\phi^3}$ &
$\lambda\sqrt{\frac{6\sigma}{\lambda}}[1+\frac{3\lambda}{2(4\pi)^2}]$  & 0\\
\hline
$\frac{d^4 V}{d\phi^4}$ &
$\lambda[1+\frac{9\lambda}{2(4\pi)^2}]$  &
$\lambda[1-\frac{3\lambda}{2(4\pi)^2}(\ln2-i \pi)]$\\
\hline
\hline
\end{tabular}
\end{center}
\end{table}

\end{document}